\documentclass[12pt]{article}
\begin{document}
\noindent Math. Model. Nat. Phenom.\\
Vol. 10, No. 5, 2015, pp. 1-5\\
DOI: 10.1051/mmnp/201510501\\

\vspace*{2cm}\centerline{\Large \bf  Three waves of chemical dynamics}

%*******************************************************************
%AUTHORS - THE CORRESPONDING AUTHOR NEEDS TO SPECIFY HIS/HER E-MAIL ADDRESS AS A FOOTNOTE
%*******************************************************************

\author{A.N. Gorban\inst{1}\thanks{\email{ag153@le.ac.uk}} \sep G.S. Yablonsky \inst{2}}

\centerline{\bf A.N. Gorban$^a$\footnote{Corresponding author. E-mail: ag153@le.ac.uk},
 G.S. Yablonsky$^b$}

\vspace{0.5cm}

\centerline{$^a$Department of Mathematics, University of Leicester, Leicester, LE1 7RH,
UK}
\vspace{0.2cm}
\centerline{$^b$Parks College of Engineering, Aviation and Technology,}

\centerline{Saint Louis University, Saint Louis, MO63103, USA}

%*******************************************************************
%ABSTRACT
%*******************************************************************

\vspace*{1cm}

\noindent {\bf Abstract.}Three epochs in development of chemical dynamics are presented. We try to understand the modern research programs in the light of classical works.

%*******************************************************************
%KEYWORDS
%*******************************************************************

The first Nobel Prize in Chemistry was awarded in 1901 to Jacobus H. van't Hoff ``in
recognition of the extraordinary services he has rendered by the discovery of the laws of
chemical dynamics and osmotic pressure in solutions''. This award celebrated the end of
the first epoch in chemical dynamics, discovery of the main laws. This epoch begun in
1864 when Waage and Guldberg published their first paper about Mass Action Law
\cite{WaageGuldberg1864}. Van't Hoff rediscovered this law independently in 1877. In 1984
the first book about chemical dynamics was published, van't Hoff's ``\'Etudes de
Dynamique chimique \cite{vantHoff1896}'', in which he proposed the `natural'
classification of simple reactions according to the number of molecules that are
simultaneously participate in the reaction. Despite his announcement, ``I have not
accepted a concept of mass action law as a theoretical foundation'', van't Hoff did the
next step to the development of the same law. He also studied the relations between
kinetics and thermodynamics and found the temperature dependence of equilibrium constant
(the van't Hoff equation). In particular, he proved the ``principle of mobile
equilibrium'': the equilibrium in a system tends to shift in such a direction as to
oppose the temperature change which is imposed upon the system. This principle was later
on generalized by Le Chatelier. The temperature dependence of the reaction rate constants
was analyzed further by Arrhenius (Nobel prize in chemistry in 1903). The famous
Arrhenius equation is a particular case of the van't Hoff equation. Chemical dynamics was
developed in line with thermodynamics. In particular, chemical equilibrium was recognised
as a dynamic process, i.e. the balance between the forward and backward reactions. Their
rates must be equal at the chemical equilibrium.  During this period, Gibbs created
modern chemical thermodynamics \cite{Gibbs1876}. The kinetic law of a single reaction and
its relation to thermodynamics were studied in full. Chemical dynamics was prepared to
the next epoch, that is to the analysis of complex reaction networks.

The scientific context of the first decades of chemical dynamics was also very important.
Maxwell \cite{Maxwell1867} and Boltzmann \cite{Boltzmann1872} developed physical kinetics
of gases. They used analogue of the Mass Action Law for collisions and discover the
principle of detailed balance: at the equilibrium, each elementary process should be
equilibrated by the reverse process. Maxwell used the principle of sufficient reason to
justify detailed balance. For the same purpose Boltzmann employed  reversibility of
collision (for more detail see \cite{GorbanResPhys2014}). Detailed balance was used by
Boltzmann to prove the $H$-theorem (entropy growth) in kinetics. Lorentz raised an
objection against this theory \cite{Lorentz1887}. He stated that the collision of
polyatomic molecules are irreversible and therefore the principle of detailed balance is
wrong. Boltzmann immediately invented a weaker condition, the principle of cyclic balance
(it is known also as semidetailed balance or complex balance) \cite{Boltzmann1887} which
is sufficient for the proof of $H$-theorem. Much later, Stueckelberg proved that the
cyclic balance condition should hold for all systems with Markov micro-description
\cite{Stueckelberg1952} (he used the $S$-matrix representation of collisions, for more
detail see \cite{GorbanShahzad2011}). Finally, in 1981, it was demonstrated that Lorentz
was wrong and detailed balance holds for collisions of polyatomic molecules
\cite{CercignaniLampis1981}. In 1902, equilibrium statistical mechanics was developed by
Gibbs and mechanical backgrounds of thermodynamics became clear \cite{Gibbs1902}.

In 1901, Wegscheider studied reaction networks which consist of several elementary
reactions \cite{Wegscheider1901}. He  discovered that the equilibrium of such a network
may not be an equilibrium of each reaction from the network if we define their rate
constants independently (``Wegscheider's paradox''). He found the conditions on the
reaction rate constants which are necessary and sufficient for coincidence of the
equilibrium of the network with the joint equilibrium of their elementary reactions
(Wegscheider's conditions of detailed balance). For example, in a simple linear cycle
these conditions state that clockwise and anticlockwise products of reaction rate
constants coincide. (The modern explanation could be found in many textbooks
\cite{Yablonskiiatal1991}.) Of course, if we define the rate of each backward reaction
thermodynamically, through the thermodynamic equilibrium constant and the reaction rate
of the forward reaction, then Wegscheider's conditions are satisfied automatically and
Wegscheider's paradox vanishes: the elementary reactions ``do not know'' about other
reactions, they all ``just know'' the same thermodynamic properties of the reagents.
Wegscheider's conditions were used by Onsager \cite{Onsager1931} in his work about
reciprocal relations awarded by Nobel Prize in Chemistry in 1968. The Onsager relations
are, in their essence, the detailed balance conditions linearized near thermodynamic
equilibrium.

Wegscheider's work was a first milestone in analysis of chemical reaction networks. The
great achievement here was the theory of chain reactions and of the critical effects in
chain reactions \cite{Semenov1935,Hinshelwood1940} In 1956, Nobel Prize in Chemistry was
awarded to Semenov and Hinshelwood ``for their researches into the mechanism of chemical
reactions''. The great problem approached by the theory of chain reaction may be
formulated as follows: {\em how the structure of reaction network affects its dynamics?}
Of course, it is solved only partially and remains the source of challenges for several
generations of researchers. It appears that the work with this problem requires methods
for model reduction \cite{Christiansen1953}. Besides the obvious reason (it is easier to
analyse systems of smaller dimension) there is a very important issue: in the whole
system the critical effects are hidden. To see the critical effect, we have to simplify
the model and to separate variables into fast and slow ones. The bifurcations in the fast
subsystem correspond to observable critical phenomena like inflammation, ignition or,
inverse, to quenching. This trick allowed Semenov and Hinshelwood to study critical
effects in chain reactions. Several methods for model reduction were developed. Most
famous of them are: quasi-steady states, quasi-equilibrium states and limiting steps and
subsystems. Later on they were integrated into several more general technologies: methods
of invariant manifolds \cite{GorKar,GorbanKarlinBAMS} and computation singular
perturbations \cite{LamGousis1994}. The mathematical backgrounds of all these methods are
the theory of singular perturbations \cite{Segel89} and the theory of invariant manifolds
of dynamical systems. Non-trivial dynamics of chemical systems attracted much attention
and various oscillations, bifurcations and non-linear waves were found. Among them the
famous example gives the Belousov-Zhabotinskii oscillatory reaction
\cite{ZaikinZhabotinsky1970}.

In 1965, the scientific program of new synthesis and rational analysis of chemical
reaction networks was proposed  by Aris \cite{Aris1965,Aris1968}. His enthusiasm,
creativity and research reputation attracted many scientists and a series of research
project was performed
\cite{HornJackson1972,Feinberg1972,Feinberg1987,Krambeck1970,OsterPerelson1974,Sellers1966}.
The Aris program was driven by the needs of chemical engineering \cite{Aris1965Reactor}.
Nevertheless, the fundamental results of this program spread wider and are now important
to various areas of science. Horn, Jackson and Feinberg rediscovered Boltzmann's cyclic
balance in the context of chemical kinetics \cite{HornJackson1972,Feinberg1972} (they
call it the ``complex balance'' condition). This complex balance condition is sufficient
for positivity of entropy production (it is the exact analogue of the 1887 version of
Boltzmann's $H$-theorem \cite{Boltzmann1887}). It guarantees decrease of the Helmholtz
free energy of closed systems under isothermal isochoric conditions and of the Gibbs free
energy (the free enthalpy) under isothermal isobaric conditions. It appears that this
condition is important not only in chemical engineering \cite{SzederkHangos2011} but also
in algebraic geometry and related areas, in particular, in the theory of toric dynamical
systems \cite{CraciunAtAl2009}. There are numerous links between chemical dynamics and
modern algebraic geometry. Classification of the limits of systems with detailed balance
when some of reactions become irreversible \cite{GorbYabCES2012} practically coincides
with the recently proved classification of binomial manifolds \cite{GrigorievMilman2012}.
The methods of tropical geometry are efficient for model reduction in large reaction
networks \cite{NoelGrigoriev2014}.

It is worth to mention other types of global Lyapunov functions which exist for some
reaction mechanisms. For example, if we consider the reaction networks with elementary
reactions of the form $mA_i \to \ldots$ (with various coefficients $m$ and right hand
sides of the reaction equations, but with a positive linear conservation law like
conservation of mass) then the $l_1$-distance between all kinetic curves decreases in
time \cite{GorBykYab1986}. For systems with arbitrary monotone kinetics the Lyapunov
functions are constructed from reaction rate functions \cite{Al-RADAngeli2014}.

The question about connection between the structure of reaction network and its dynamics,
remains one of the central problem of chemical dynamics. In parallel with answering this
question we need to answer the question what is a kinetic law of complex chemical
reaction. If we separate times and select fast and slow subsystems then this second
question could be understood as an elimination problem: what are equations of slow
dynamics after elimination of fast variables? This problem stimulated development of a
new chapter of computer algebra \cite{BykovKytmanovLazman}. For several classes of
catalytic reactions this exclusion was performed and the steady-state kinetic law of the
single overall reaction was presented as the single polynomial regarding the reaction
rate. Such a polynomial may have several roots which correspond to different
steady-states. At the same time its free term has a rigorous thermodynamic form and
validates thermodynamic correctness of this presentation \cite{LazYab2008}.

Various methods of graph theory were employed for analysis of chemical reaction networks.
The general theory of equations on graphs was developed \cite{VolpertKhudyaev1985}. Now,
several areas of applied and pure mathematics are used in chemical dynamics and are
developed further due to this applications. ``The beginning of this era was marked by the
concerted effort of a few to raise the mathematical consciousness of the profession to
think fundamentally about processes'' \cite{RamkrishnaAmundson2004}. After these 50 years
of efforts we have to look back, to review the past and to ask about the future: what
should we expect soon? New technologies will generate new questions. The quantum world
will become closer to industry. Biological engineering will be as usual as is chemical
engineering. The artificial intelligence and rapid computation will change the practice
of mathematical modelling.

Three eras (or waves) of chemical dynamics can be revealed in the flux of research and
publications. These waves may be associated with leaders: the first is the van't Hoof
wave, the second may be called the Semenov--Hinshelwood wave and the third is definitely
the Aris wave. Of course, the whole building was impossible without efforts of hundreds
of other researchers. Few of of them are mentioned in our brief review, more are cited in
the papers of the issue.

This issue is opened by  the reprint of the classical paper of A.I. Volpert, where he
introduced the differential equations on graphs. The work of Gorban and Kolokoltsov
analyzed appearance of Mass Action Law with complex balance conditions and their
generalizations in the Michaelis--Menten--Stueckelberg limit of general Markov processes.
Joshi and Shiu studied minimal reaction networks with multiple steady states. In the
paper by Bykov and Tsybenova  the classical nonlinear models of catalytic reaction were
augmented by the additional variable, i.e. temperature, and the extended models were
systematically studied. M\"uller and Hofbauer applied the formalism of chemical reaction
networks to kinetics of genetic recombination and analyze existence, uniqueness, and
global stability of an equilibrium in such networks. Grigoriev, Samal, Vakulenko and
Weber developed algebraic algorithms for analysis and reduction of larger metabolic
reaction networks and studied several biochemical networks. Constales, Yablonsky and
Marin analyzed kinetics of pulse-response experiments in the Temporal Analysis of
Products (TAP) setup and demonstrated that in these special conditions the activity
profile of a prepared catalytic system depends only on the total amount of admitted
substance. Gorban demonstrated how to use a special geometric procedure,
forward-invariant peeling, to produce forward-invariant subset from a given set in
concentration space space and to prove persistence of a chemical dynamic system.


\begin{thebibliography}{99}

\bibitem{Al-RADAngeli2014}M.A. Al-Radhawi, D. Angeli. {\em Robust Lyapunov functions for
    complex reaction networks: An uncertain system framework.} In Decision and Control (CDC), 2014 IEEE
    53rd Annual Conference, IEEE, 2014, 3101--3106.

\bibitem{Aris1965Reactor}R. Aris. Introduction to the Analysis of Chemical Reactors, Prentice Hall,
Englewood Cliffs, NJ, 1965.

\bibitem{Aris1965}R. Aris. {\em Prolegomena to the rational analysis of systems of
    chemical reactions}. Arch. Ration.  Mech. Anal., 19 (2) (1965), 81--99.

\bibitem{Aris1968}R. Aris. {\em Prolegomena to the rational analysis of systems of
    chemical reactions II. Some addenda}. Arch. Ration.  Mech. Anal.,
    27 (5) (1968), 356--364.

\bibitem{Boltzmann1872}L. Boltzmann, {\em Weitere Studien \"uber das
    W\"armegleichgewicht unter Gasmolek\"ulen}, {Sitzungsberichte der Kaiserlichen Akademie der Wissenschaften in
    Wien}, {66} (1872),
    275--370.

\bibitem{Boltzmann1887}{L. Boltzmann}. {\em Neuer Beweis zweier S\"atze \"uber das
    W\"armegleichgewicht unter mehratomigen Gasmolek\"ulen}. {Sitzungsberichte der
    Kaiserlichen Akademie der Wissenschaften in
    Wien}, {95} (2) (1887), {153--164}.

\bibitem{BykovKytmanovLazman}V.I. Bykov, A.M. Kytmanov, M.Z. Lazman. Elimination methods
    in polynomial computer algebra. Mathematics and its Applications, V. 448. Springer,
    1998.

\bibitem{CercignaniLampis1981}C. Cercignani, M. Lampis. {\em On the $H$-theorem for
    polyatomic gases}. J. Stat. Phys., 26 (4) (1981) 795--801.

\bibitem{Christiansen1953}J.A. Christiansen.  {\em The elucidation
    of reaction mechanisms by the method of intermediates in
    quasi-stationary concentrations.} { Adv. Catal.} , { 5} (1953),
    311--353.

\bibitem{CraciunAtAl2009}G. Craciun, A. Dickenstein, A. Shiu, B. Sturmfels. {\em Toric
    dynamical systems.} J. Symb. Comput., 44 (11) (2009), 1551--1565.

\bibitem{GrigorievMilman2012}D. Grigoriev, P.D. Milman. {\em Nash resolution for binomial
    varieties as    Euclidean   division. A priori termination bound, polynomial complexity in essential dimension 2.}
    Advances in Mathematics, 231 (6) (2012), 3389--3428.

\bibitem{Feinberg1972}M. Feinberg. {\em Complex balancing in general kinetic systems.}
    Arch. Ration.  Mech. Anal., 49 (1972), 187--194.

\bibitem{Feinberg1987}Feinberg, M. {\em Chemical reaction network structure and the
    stability of complex isothermal reactors: I. The deficiency zero and deficiency one
    theorems}. Chem. Eng. Sci., 42 (10) (1987), 2229--2268.

\bibitem{Gibbs1876}J.W. Gibbs. {\em On the equilibrium of heterogeneous substances.}
    Trans. Conn. Acad. Art. Sci., 3 (1876-1878), 108--248, 343--524.

\bibitem{Gibbs1902}J.W. Gibbs. Elementary Principles in Statistical Mechanics, developed
    with especial reference to the rational foundation of thermodynamics. Yale
    Bicentennial Publications. Scribner and Sons, NY, 1902. [Dover Publications Inc.;
    Reprint edition, 2015.]

\bibitem{GorbanResPhys2014}A.N. Gorban. {\em Detailed balance in micro- and macrokinetics
    and micro-distinguishability of macro-processes}. Results in Physics 4 (2014),
    142--147.

\bibitem{GorBykYab1986}A.N. Gorban,  V.I. Bykov,  G.S. Yablonskii. {\em Thermodynamic
    function analogue for reactions proceeding without interaction of various substances.} Chem. Eng. Sci.,
    41 (11) (1986), 2739--2745.

\bibitem{GorKar}A.N. Gorban, I. Karlin. Invariant Manifolds for Physical and Chemical Kinetics (Lecture
    Notes in Physics).  Springer, 2005.

\bibitem{GorbanKarlinBAMS}A.N. Gorban, I. Karlin. {\em Hilbert's 6th Problem: exact and
    approximate hydrodynamic manifolds for kinetic equations.} Bulletin of the American
    Mathematical Society, 51(2) (2014), 186--246.

\bibitem{GorbanShahzad2011}A.N. Gorban, M. Shahzad. {\em The
    Michaelis--Menten--Stueckelberg
    Theorem.} {Entropy}, {13} (2011) {966--1019}. Corrected postprint: arXiv:1008.3296.

\bibitem{GorbYabCES2012}A.N. Gorban, G.S. Yablonskii. {\em Extended detailed balance for
    systems with irreversible reactions.} Chem. Eng. Sci., 66 (2011) 5388--5399. arXiv:1101.5280
    [cond-mat.stat-mech].

\bibitem{Hinshelwood1940}C.N. Hinshelwood. The Kinetics of Chemical Change. The
    Clarendon press, Oxford,  1940.

\bibitem{HornJackson1972}F. Horn, R. Jackson. {\em General mass
    action kinetics.} {Arch. Ration. Mech. Anal.},
    {47} (1972), 81--116.

\bibitem{Ko10}V.N. Kolokoltsov. {Nonlinear Markov processes and kinetic equations}.
    Cambridge Tracks in Mathematics 182, Cambridge Univ. Press, 2010.

\bibitem{Krambeck1970}F.J. Krambeck. {\em The mathematical structure of chemical kinetics
    in homogeneous single-phase systems.} Arch. Ration. Mech. Anal., 38 (5) (1970), 317--347.


\bibitem{LazYab2008}M. Lazman, G. Yablonsky. {\em Overall Reaction Rate Equation of
    Single Route Catalytic Reaction},  Advances in Chemical Engineering, 34 (2008),
    47--102.

\bibitem{Lewis1925}G.N. Lewis. {\em A new principle of equilibrium}. Proceedings of the
    National Academy of Sciences, 11 (1925), 179--183.


\bibitem{Maxwell1867}J.C. Maxwell. {\em On the dynamical theory of gases.}
    Philosophical Transactions of the Royal Society of London, 157 (1867), 49--88.

\bibitem{LamGousis1994}S.H. Lam,  D.A. Goussis. {\em The CSP method for simplifying kinetics.}
    International Journal of Chemical Kinetics, 26 (4) (1994), 461--486.

\bibitem{Lorentz1887}H.-A. Lorentz. {\em \"Uber das Gleichgewicht der lebendigen Kraft
    unter
    Gasmolek\"ulen.} Sitzungsberichte der Kaiserlichen Akademie der Wissenschaften in
    Wien, 95 (2) (1887), 115--152.

\bibitem{NoelGrigoriev2014}V. Noel, D. Grigoriev, S. Vakulenko, O. Radulescu. {\em
    Tropicalization and tropical equilibration of chemical reactions.} Tropical and
    Idempotent Mathematics and Applications, Contemporary Mathematics 616 (2014),
    261--277.

\bibitem{Onsager1931}L. Onsager. {\em Reciprocal relations in irreversible processes. I},
    Phys.  Rev., 37 (1931), 405--426.

\bibitem{OsterPerelson1974}G.F. Oster, A.S. Perelson. {\em Chemical reaction dynamics.}
    Arch. Ration.  Mech. Anal., 55 (3) (1974), 230--274.

\bibitem{RamkrishnaAmundson2004}D. Ramkrishna, N.R. Amundson. {\em Mathematics in
    Chemical Engineering: A 50 Year Introspection.} AIChE Journal, 50 (1) (2004), 7--23.

\bibitem{Segel89}L.A. Segel, M. Slemrod.   {\em The     quasi-steady-state assumption: A
    case study in  perturbation.} {SIAM Rev.},  {31} (1989), 446--477.

\bibitem{Sellers1966}P.H. Sellers. {\em Algebraic complexes applied to chemistry.} PNAS U.S.A., 55 (4) (1966), 693--698.

\bibitem{Semenov1935}N.N. Semenov. Chemical kinetics and chain reactions. The Clarendon
    press, Oxford,  1935.

\bibitem{Stueckelberg1952}E.C.G. Stueckelberg. {\em Th\'eor\`eme $H$ et
    unitarit\'e de $S$.} {Helv. Phys. Acta},  {25} (1952), 577--580.

\bibitem{SzederkHangos2011}G. Szederk\'enyi, K.M. Hangos. {\em Finding complex balanced
    and detailed balanced realizations of chemical reaction networks.} J. Math. Chem.,
    49, (6) (2011), 1163--1179.

\bibitem{vantHoff1896}J. H. van't Hoff. Studies in Chemical Dynamics, F. Muller \& Co,
    Amsterdam, 1896 [A revised edition of Hoff's ``Etudes de dynamique chimique''. Revised and enlarged
    by E. Cohen, translated by T. Ewan.]

\bibitem{VolpertKhudyaev1985}A.I. Volpert, S.I.
    Khudyaev. Analysis
    in classes of discontinuous functions and equations of
    mathematical physics. Nijoff, Dordrecht, The Netherlands, 1985.

\bibitem{WaageGuldberg1864}P. Waage, C. M. Guldberg. {\em Studies concerning affinity.}
Forhandlinger: Videnskabs - Selskabet i Christinia (Norwegian Academy of Science and
Letters), (1864), 35--45. [English translation: J. Chem. Educ., 1986, 63 (12),
1044--1047.]

\bibitem{Wegscheider1901}R. Wegscheider. {\em \"Uber simultane Gleichgewichte und die
    Beziehungen zwischen Thermodynamik und Reactionskinetik homogener Systeme.}
    Monatshefte f\"ur Chemie / Chemical Monthly, 32 (8) (1901), 849--906.


\bibitem{Yablonskiiatal1991}G.S. Yablonskii, V.I. Bykov,  A.N. Gorban, V.I. Elokhin.
    Kinetic Models of Catalytic Reactions. Elsevier, Amsterdam, The Netherlands, 1991.

\bibitem{ZaikinZhabotinsky1970}A.N. Zaikin, A.M. Zhabotinsky. {\em Concentration wave
    propagation in two-dimensional liquid-phase self-oscillating system.} Nature, 225 (1970),
    535--537.


\end{thebibliography}
\end{document}